\documentclass[12pt,epsfig]{article}
\usepackage{graphicx,amsmath,amssymb}

\newlength{\dinwidth}
\newlength{\dinmargin}
\def\lapproxeq{\lower .7ex\hbox{$\;\stackrel{\textstyle
<}{\sim}\;$}}
\def\gapproxeq{\lower .7ex\hbox{$\;\stackrel{\textstyle
>}{\sim}\;$}}
\def\gtrsim{\lower .7ex\hbox{$\;\stackrel{\textstyle
>}{\sim}\;$}}
\def\lesim{\lower .7ex\hbox{$\;\stackrel{\textstyle
<}{\sim}\;$}}

\def\be{\begin{equation}}
\def\ee{\end{equation}}
\def\bea{\begin{eqnarray}}
\def\eea{\end{eqnarray}}

\def\Im{{\rm Im}}
\def\tot{{\rm tot}}
\def\inel{{\rm inel}}
\def\el{{\rm el}}
\def\gap{{\rm gap}}
\def\D{{\rm D}}

\begin{document}
\begin{flushright}
IPPP/05/24 \\
DCPT/05/48 \\
7th July 2005 \\

\end{flushright}

\vspace*{0.5cm}

\begin{center}
{\Large \bf The partonic interpretation of reggeon theory models}

\vspace*{1cm}
\textsc{K.G. Boreskov$^a$, A.B.~Kaidalov$^{a,b}$, V.A.~Khoze$^{b,c}$, \\
A.D. Martin$^b$ and M.G. Ryskin$^{b,c}$} \\

\vspace*{0.5cm}
$^a$ Institute of Theoretical and Experimental Physics, Moscow, 117259, Russia\\
$^b$ Department of Physics and Institute for
Particle Physics Phenomenology, \\
University of Durham, DH1 3LE, UK \\
$^c$ Petersburg Nuclear Physics Institute, Gatchina,
St.~Petersburg, 188300, Russia \\

\end{center}

\vspace*{0.5cm}

\begin{abstract}
We review a physical content of the two simplest models of reggeon
field theory: namely the eikonal and the Schwimmer models.
The AGK cutting rules are used to obtain the inclusive, the inelastic
and the diffractive cross sections. The system of nonlinear equations
for these cross sections is written down and analytic expressions
for its solution are obtained.
We derive the rapidity gap dependence of the differential cross sections
for diffractive dissociation in the Schwimmer model and in its eikonalized extension.
The results are interpreted from the partonic viewpoint of the interaction at high energies.
\end{abstract}

\section{Introduction}

Regge theory is widely used to describe the low $p_T$ high-energy interactions of hadrons, nuclei
and (real and virtual) photons. The theory takes into account both Regge poles and cuts.
The latter are related to the exchange of several reggeons in the $t$-channel.
The status of this theory within QCD is reviewed, for example, in Refs.~\cite{abk}.
The Pomeranchuk singularities (that is the Pomeron pole and the corresponding cuts)
play a special role in this theory as they determine the high energy behaviour of
diffractive processes and multiparticle production \cite{abk}. It is important
to understand the connection between the general results of reggeon theory and the
space-time picture of hadronic interactions.
This becomes possible due to the relation between regge theory and the parton model
\cite{fey,vng}.
Multiple pomeron exchanges are especially important if the intercept of the pomeron,
$\alpha_P(0)$, is larger than unity, that is $\Delta \equiv \alpha_P(0)-1>0$.
This so-called ``supercritical'' theory is favoured both by experimental data and
by calculations in QCD perturbation theory \cite{bfkl}. In this case the partonic
interpretation becomes very non-trivial. The relation between the probabilistic
partonic picture of the interaction and diagrams of reggeon theory has been studied
in Refs.~\cite{grass,kgb}.

In this paper we discuss two simple analytic models of regge theory, which
provide particular examples of
the partonic picture of high energy hadronic collisions.
These are the eikonal model and the Schwimmer model \cite{sch}
\footnote{~The currently popular Balitski-Kovchegov equation \cite{bk} is, from the partonic
and space-time viewpoints, a generalization of the Schwimmer model.}%
, which are often used in phenomenological applications of regge theory.
We use the AGK cutting rules \cite{agk} to obtain the inelastic, diffractive and
inclusive cross sections predicted by these models;
and discuss the partonic interpretation of these results.
Although some of these are known, it is informative to summarize them here.
For the Schwimmer model, and its eikonal generalization, we obtain explicit formulae
for the total, inelastic and diffractive cross sections. We also obtain the dependence
of the differential cross section on the size of the rapidity gap.

Our ultimate goal is to use these results to improve the `global' analysis of data for
`soft' high energy processes, see, for example, Ref.~\cite{KMRsoft}.

\section{Multiparticle content of reggeon diagrams}

An interpretation of reggeon diagrams in terms of their inner multiparticle
structure was given in Ref.\cite{afs}. It corresponds to the qualitative picture \cite{fey,vng}
that a fast hadron of momentum $p$ (of rapidity $y\simeq\ln 2p/m$) interacts with a target
due to quantum-mechanical fluctuations containing slow particles.
The structure of the fluctuations is rather specific and is usually called
`multiperipheral'.
Such a fluctuation contains $\sim\ln p$ soft virtual particles ordered in their
rapidities.
For brevity we shall call these particles {\em soft partons}, or simply {\em partons}
\footnote{~Here we do not associate (soft) {\em partons} with definite objects like quarks,
gluons, or pions, because only rather general features of them \cite{fey,vng} are relevant
for our analysis. Hard partons of different spatial scales are considered in connection with
deep inelastic scattering and other hard processes.}%
.
Only the slowest partons
\footnote{~These partons have momenta of the order of the typical hadronic scale $\mu$
of about several hundred MeV.}
have a chance to interact directly with a target. The faster partons of the fluctuation
simply play the role of spectators.
The cross section of the interaction is proportional to the number $n(y)$ of slow partons.
In this scheme, slow partons originate from faster partons close
in rapidity and $n(y)$ has the exponential behaviour $\sim \exp(\Delta y)$.
The interaction of a single slow parton corresponds to regge-pole behaviour of cross section
with $\Delta=\alpha_P(0) - 1$,
while the interactions of two or more partons with the target give rise to regge-cut-type
contributions.
If the (multiperipheral) evolution of a fast parton into slow ones is independent
of the evolution of the other fast partons, then we obtain independent slow partons
whose interactions correspond to 'non-enhanced' reggeon diagrams of the eikonal approximation
(see section 3).

For the supercritical pomeron, i.e. described by a regge pole with $\Delta > 0$,
the number of slow partons increases exponentially with the initial rapidity $y$,
i.e. in the course of the evolution of the parton fluctuation in rapidity space
the number of partons multiplies, for example, by a splitting mechanism.
As a consequence another type of reggeon diagrams will appear -- that is
`enhanced' diagrams of the Schwimmer type occur (see section 4).
We emphasize that for the parton dynamics to be consistent we require
not only splitting, but also fusion of partons -- though in special cases, we may,
to a good approximation, neglect the latter process.

The Abramovsky-Gribov-Kancheli (AGK) cutting rules \cite{agk} are a powerful tool
for the investigation of the multiparticle structure of complicated reggeon
diagrams.
They were derived as the high-energy version of Cutkosky cutting rules \cite{cut}.
They give the discontinuity of the whole reggeon diagram in terms of the discontinuities
of its component subdiagrams. Each reggeon diagram has various discontinuities
which correspond to different ways of cutting the diagram and to different intermediate states.
For example, cutting the regge pole diagram corresponds to the simple multiperipheral
intermediate state. On the other hand, cutting a double-pomeron-exchange diagram
leads to intermediate states of both double and single density,
depending on the number of cut pomerons, and also on the state with a large rapidity gap
obtained when the diagram is cut {\em between} pomerons.
The AGK rules give relations between the contributions of given reggeon diagrams
to different multiparticle cross sections. Examples of such relations
will be discussed below.

The space-time picture of the interaction is another valuable tool in the description
of high energy collisions.
Pomeron exchange is a highly non-local process. It is characterised by longitudinal
and time scales which are proportional to the initial energy. As a consequence,
only reggeon diagrams with so called {\em non-planar} structure contribute at high energies.
Partonic fluctuations for these diagrams develop simultaneously at {\em comparable}
longitudinal distances.
In contrast, fluctuations corresponding to {\em planar} diagrams develop {\em succesively}
and will only contribute for a very extended target. However, particular discontinuities
of planar graphs (which vanish when summed) can be asymptotically essential, and must
therefore be taken into account in the analyses of cross sections of particular processes.
For instance, the elastic cross section is evidently determined by cutting a planar diagram.

Gribov \cite{vng} managed to present the total cross section in a very simple way
through dispersion integrals of the `particle-particle $\rightarrow n$~pomeron' amplitudes.
He did this by rearranging the contributions of both non-planar and planar
multireggeon diagrams using the reggeon unitarity condition.
Keeping only the one-particle pole contributions to these amplitudes (Fig.~\ref{fig:1}(a))
we reproduce the formula of the well-known `eikonal approximation' for high-energy scattering.
However we should keep in mind that the space-time picture behind this formula does not
correspond to successive elastic rescatterings.
The genuine space-time picture has been lost under rearrangement of diagrams
with different planarities.

\section{The eikonal model}

\subsection{The eikonal $\chi_P$}
The single pomeron-exchange amplitude has the form
\footnote{~The normalization of amplitude is $\sigma^{\tot}(s)=2\Im M(s,0)$.}
\begin{align}
\label{eq:fteik}
M_P(s,t) = \left(\frac{s}{s_0}\right)^{\alpha_P(t)-1} \eta_P(\alpha_P(t))\;g_1(t)g_2(t) ~,
\end{align}
where $g_{1,2}(t)$ are the couplings of the pomeron to the colliding hadrons, and
\begin{align}
\eta_P(\alpha_P(t)) = -\frac{1+\exp(-i\pi\alpha_P(t))}{\sin\pi\alpha_P(t)}
\end{align}
is the signature factor, which for $\alpha_P(0)=1$ is equal to $i$.
Here we shall neglect the real part of the pomeron amplitude,
assuming that the value of $\Delta = 1 - \alpha_P(0)$ is small.

It is convenient to analyse high-energy interactions in terms of the impact parameter, $b$,
by introducing the Fourier transformed amplitude
\footnote{~The normalization used for $M(s,t)$ corresponds to
$\sigma^{\rm tot}=2\int d^2 b ~{\rm Im}f(s,b)$ and
$\sigma^{\rm el}=\int d^2 b~ |f(s,b)|^2$.}%
\begin{align}
\label{Fourier}
f(s,b) = \frac{1}{(2\pi)^2}\int d^2 k\, e^{i \mathbf{b\cdot k}}\, M(s,\mathbf{k}^2) ~,
\end{align}
where $t=-\mathbf{k}^2$.
For an exponential parametrization of residues, $g_i(t) = g_i \exp(-R_i^2 \mathbf{k}^2)$,
$i = 1,2$,
and a linear parametrization of the pomeron trajectory,
$\alpha_P(t)=\alpha_P(0)+\alpha^\prime _P t$,
we obtain the familiar regge-pole approximation of the amplitude in impact parameter space
\begin{align}
\label{pole_b}
& f_P(Y,b) \simeq i\, g_1 g_2\, \frac{\exp\left(-\dfrac{b^2}{4\lambda}\right)}{4\pi\lambda}\exp(\Delta Y)
 = i\,\frac{\chi_P (Y,b)}{2}  ~, \\[2mm]
& Y\simeq \ln(s/s_0) ~, \quad s_0 \simeq 1\text{ GeV}^2 ~, \nonumber\\[2mm]
& \lambda=R_1^2+R_2^2+\alpha_P' Y ~. \nonumber
\end{align}
Note that the amplitude $f_P(Y,b)$ increases as $\exp(\Delta Y)$, and so violates unitarity
as $s \to \infty$.
(Recall that $|f(Y,b)| \leq 2$ due to unitarity.) If we include multi-pomeron exchanges,
then this inconsistency is avoided.

\subsection{Cross section formulae in terms of the eikonal $\chi_P$}
The eikonal approximation is the simplest way to restore $s$-channel unitarity
for elastic amplitudes.
Summation of the eikonal diagrams gives the following well known expressions
in impact parameter space
\begin{align}
\label{eq:eik}
f(Y,b) &= i\,\left(1- \exp\left(-\frac{\chi_P (Y,b)}{2}\right)\right) ~, \\[2mm]
\label{eik_tot}
\sigma^{\rm tot}(Y) & = 2\int d^2b\;\left( 1-\exp\left(-\frac{\chi_P (Y,b)}{2}\right)\right) ~,
\end{align}
where, recall, $Y\simeq \ln (s/s_0)$.
At very high energies, $\Im f(Y,b) \to 1$ (the black disc limit) in the region
of $b$ where $\chi_P(Y,b)$ is large. From \eqref{pole_b} we see that $\chi_P$ becomes small
only in the region $b^2 > 4\,\Delta\alpha' Y^2$.
Thus, for very large $s$, the radius of interaction increases as
$R^2(s)=4\,\Delta\alpha'\ln^2(s/s_0)$, and the total cross section increases as
$\sigma^{\rm tot} \approx 2\pi R^2 (s)$.

\begin{figure}
\begin{center}
\includegraphics[width=0.9\textwidth]{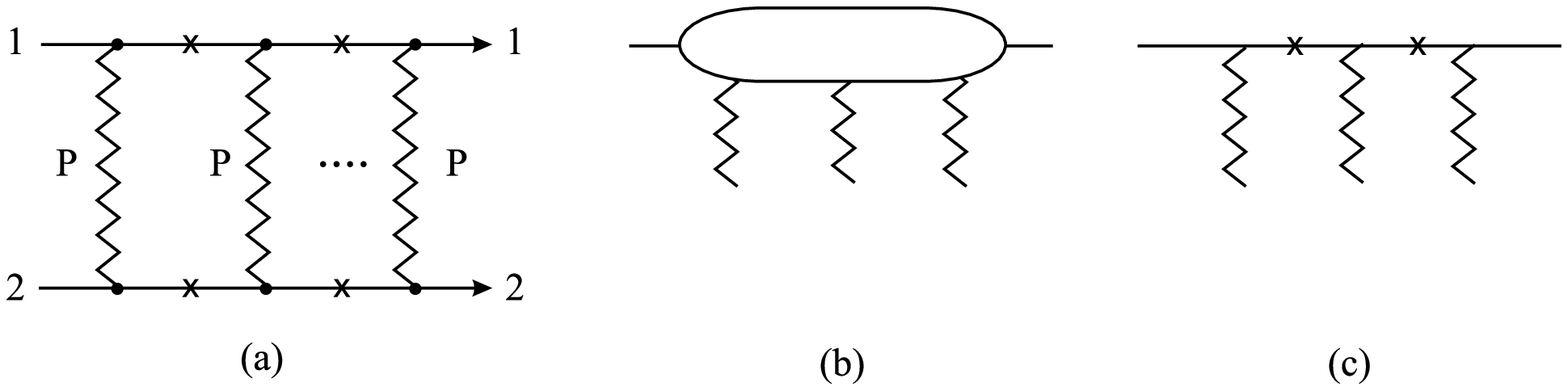}
\caption{(a) The diagrams of the eikonal model in Regge theory;
in which the \mbox{2-particle $\to$} \mbox{$n$-pomeron} amplitude of diagram (b) is represented
by diagram (c). A cross on a line means that in the expression for the discontinuity
of the amplitude of diagram (a) the propagator $1/(q^2-m^2)$ is replaced by
$2\pi i \delta (q^2-m^2)$.}
\label{fig:1}
\end{center}
\end{figure}
%



To obtain the inelastic cross section we must consider, according to the AGK cutting rules,
all eikonal type diagrams in which at least one pomeron is cut. Then for each cut pomeron
we have a factor $\chi_P$, and for each uncut pomeron a factor $(-\chi_P)$, which takes
into account the position of the uncut pomerons both on the left and on the right
of the cutting plane (that is $if_P-if_P^*=-\chi_P$).
If no pomerons are cut, then it is necessary to subtract the extra terms where all pomerons
formally are on the same side of the cutting plane.
This rule is valid both in the momentum and in the coordinate representation \cite{bkai}.

For instance, for the two-pomeron-exchange diagram, the discontinuities for zero,
one and two cut pomerons give, respectively,
\begin{align}
\label{cut2}
\sigma_0^{(2)} = \frac{1}{2!}\left[(-\chi_P)^2 -2 (-\chi_P/2)^2 \right]
= \frac{1}{4}\chi_P^2 ~, ~
\sigma_1^{(2)} = \frac{1}{2!} 2\chi_P (-\chi_P) = -\chi_P^2 ~, ~
\sigma_2^{(2)}(b) = \frac{\chi_P^2}{2!} ~,
\end{align}
which reproduces the known AGK ratios $1:-4 : 2$ \cite{agk}. It is easy to check that
these contributions sum to the second term, $2\,\Im f^{(2)}(b)$,
in the power series expansion of the eikonal formula~\eqref{eq:eik}.

In this model, the distribution in terms of the number $k$
of cut pomerons at fixed $b$ has the Poissonian form
\begin{align}
\label{eq:inel_k}
\sigma_k(Y,b) & = \frac{(\chi_P(Y,b))^k}{k!}~\exp(-\chi_P(Y,b)) ~,\\
\sigma_0(Y,b) & = 1 + \exp(-\chi_P(Y,b)) - 2 \exp\left(-\frac{\chi_P(Y,b)}{2}\right) ~,
\label{eq:0}
\end{align}
which leads to the following expressions for the inelastic and diffractive cross sections
\begin{eqnarray}
\label{eik_inel}
\sigma^{\inel}(Y)~ & = &~\int d^2b~\left( 1-\exp(-\chi_P(Y,b))\right) \\
\sigma_0(Y)~& = & ~\int d^2b~\left( 1-\exp\left(-\frac{\chi_P(Y,b)}{2}\right)\right)^2.
\label{eik_diff}
\end{eqnarray}
In the eikonal model of Fig.~\ref{fig:1} only elastic intermediate states appear in the
rescattering diagrams%
~\footnote{~For the interaction with a nucleus, the AGK rules can be applied
in a more general situation in which every pomeron exchange is substituted
by the hadron-nucleon amplitude $f_{hN}(Y,b)$, i.e. by the whole set of multi-pomeron exchanges.
The amplitudes $f_{hN}(Y,b)$ have both inelastic and elastic discontinuities.
As a result, \eqref{eik_tot} will contain $\sigma_{hN}^{\tot}(Y,b)/2$ instead of $\chi_P/2$,
\eqref{eik_inel} will contain $\sigma_{hN}^{\inel}(Y,b)$ and \eqref{eik_diff} will
include both the elastic and diffractive dissociation cross sections. \cite{bkai}}%
\label{f1}%
. Hence it is natural that $\sigma_0=\sigma^{\rm el}$.
\begin{figure}
\begin{center}
\includegraphics[width=0.3\textwidth]{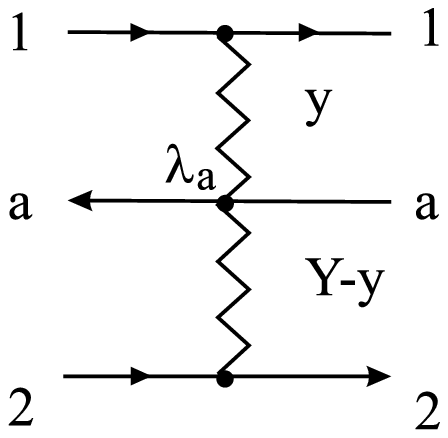}
\caption[The Mueller]{The Kancheli-Mueller diagram \cite{ovk,ahm} describing the single particle inclusive
process $12 \to aX$ in the eikonal model.  $y$ is the rapidity of particle $a$, and $Y$ is the
rapidity interval between the colliding hadrons.}
\label{fig:2}
\end{center}
\end{figure}

The AGK cancellation theorem \cite{agk} enables the inclusive cross sections to be calculated.
For example, consider the single particle inclusive process $12 \to aX$. In this case, at least
one pomeron is cut. The others may be either cut (giving a contribution $\chi_P$) or uncut
(giving a contribution $-\chi_P$). Thus the
multiple-pomeron-exchange contributions cancel, and the single-inclusive cross section
is described by the diagram shown in Fig.~\ref{fig:2}.  As a function
of the particle rapidity it is given by \be \frac{d\sigma^a}{dy} ~ =
~\lambda_a g_1(0)g_2(0) e^{\Delta y} e^{\Delta(Y-y)} ~ = ~\lambda_a
g_1(0)g_2(0) e^{\Delta Y}, \label{eq:aincl} \ee where $\lambda_a$ is
related to the rapidity density of hadron $a$ in events originating
from single-pomeron-exchange.

Note that the impact parameter $b$ is conserved during the eikonal rescattering.
As a consequence, all the formulae are valid, not only for integrated cross sections,
but at any fixed $b$.
In particular, the same increase with energy, $~(s/s_0)^\Delta$, occurs at fixed $b$
for the density of particles $(d\sigma^a/dy)/\sigma^{\inel}$ in the limit $s \to \infty$.

\subsection{Partonic interpretation of the eikonal model}
\label{partons_eik}

\begin{figure}
\begin{center}
\includegraphics[height=5cm]{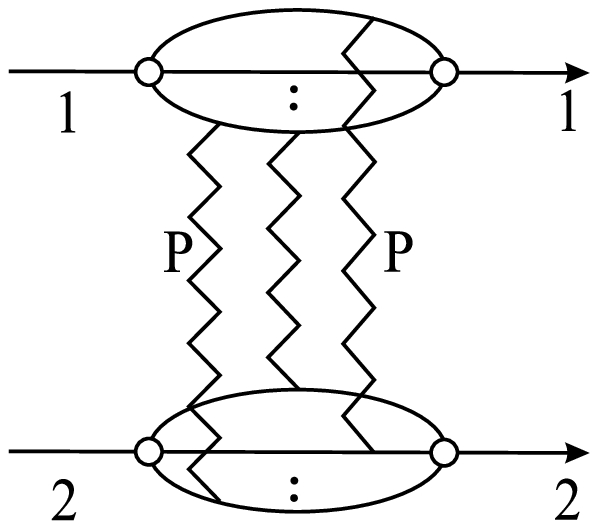}
\caption{The Feynman diagrams for the interaction of partonic fluctuations of colliding
hadrons.\label{fig:3}}
\end{center}
\end{figure}

The formulae of the eikonal approximation can be interpreted in terms of the interactions
of {\it fast} partons in colliding hadrons.
If the distribution in the number of fast partons in a hadron has a Poisson form, and if
the
cross section of a parton-parton interaction is denoted by $\widehat{\sigma}$,
then the summation of the diagrams
\footnote{~Note that in these diagrams a parton of hadron 1 interacts with only
one parton of hadron 2. It resembles the Czy\'{z} - Maximon approximation \cite{cm}
for nucleus-nucleus collisions, when only single nucleon-nucleon interactions are
taken into account.}
shown in Fig.~\ref{fig:3} leads to the eikonal results of (\ref{eq:eik}) and (\ref{eik_inel})
with
\begin{align}
\chi_P(Y,b)&=\int d^2b_1 d^2b_2\,\rho_{1}^{(0)} (\mathbf{b}_1)\;
\widehat{\sigma}(Y,\mathbf{b}-\mathbf{b}_1+\mathbf{b}_2)\;
\rho_{2}^{(0)}(\mathbf{b}_2) ~,
\label{eq:partoneik}
\end{align}
where $\rho_{i}^{(0)}(b_i)$ is the fast parton distribution
\footnote{~The distributions are normalized so that
$\int d^2b_i~\rho_{i}^{(0)}(b_i)=\langle n_{i} \rangle$, where
$\langle n_{i} \rangle$ is the mean number of fast partons
in the hadron $i$.}
in impact parameter space of colliding hadron $i$, with $i=1,2$.
That is the contribution of single-pomeron-exchange is represented by the single particle
densities $\rho_{i}^{(0)}$ and the
cross section $\widehat{\sigma}(Y)$ of the parton-parton interaction.
Similarly, the $n$-pomeron-exchange contribution equals to the probability of finding
$n$ partons in each of the colliding hadrons (which, for independently distributed partons,
are simply $(\rho_{i}^{(0)})^n$ with $i=1,2$) multiplied by a sign-alternating factor
$(-1)^{n+1}(\widehat{\sigma} /2)^n/n!$. The alternating sign is due to parton screening.

Note that (\ref{eq:partoneik}) is written assuming that the partons are all the same.
In the more realistic situation, with quarks and gluons as partons, the formalism may be more
complicated to allow for different parton-parton amplitudes. It results in a straightforward
generalization of the simple eikonal model, see, for example Ref.~\cite{KKMR}.

One of the drawbacks of this simple model is the lack of energy-momentum
conservation \cite{cap}. Indeed a very large number of interactions
$\approx \chi_P(Y)$ become important as $Y \to \infty$, and it is necessary to allow
for energy-momentum conservation in the distributions of momenta
in the partonic systems.
This will lead to deviations from Poisson distributions. These deviations are usually taken
into account in realistic models of high-energy interactions \cite{abk,cap}.

However, the interpretation is not self-consistent in the case of the supercritical pomeron
($\Delta > 0$).
We see that in this case the origin of the increase of the amplitude
$\widehat{f}(Y)$ with energy is ${\it not}$ explained in terms of partons.
It is desirable to reformulate this approach without reference to pomeron exchange
in parton-parton interaction.
As we discussed in section 2, the increase of the pomeron-exchange amplitude can be
explained as the increase in the number of {\it slow} partons.
It is possible to rewrite (\ref{eq:partoneik}) in such form  that it will correspond
to the interaction of two partonic showers viewed from the Lorentz frame at arbitrary
rapidity $y_1$:
\be
\chi_P(Y,b)=\int d^2b_1 d^2b_2\,\rho_{1} (y_1,\mathbf{b}_1)\;
\widehat{\sigma}_0(\mathbf{b}-\mathbf{b}_1+\mathbf{b}_2)\; \rho_{2} (Y-y_1,\mathbf{b}_2) ~,
\label{eq:partonYeik}
\ee
where the slow parton densities $\rho_{i}$ have a Regge form
\begin{align}
\label{green}
\rho_{i}(y,b) = \frac{g_i}{4\pi\alpha' y}\exp\left(-\frac{b^2}{4\alpha' y}\right)\,
\exp(\Delta y) ~,
\end{align}
and the parton-parton interaction
cross section $\widehat{\sigma}_0$ is local in rapidity space.
It is easy to see that expression (\ref{eq:partonYeik}) does not depend on
a choice of Lorentz frame, i.e. on the point $y_1$, due to particular form
of reggeon densities \eqref{green}.

Since only the products of quantities occur, we have been able to move the energy dependence
(that is the $s$ or $Y$ dependence) from the parton-parton reaction
cross section $\widehat{\sigma}(Y)$ to the incoming parton distributions $\rho_{i}$ in \eqref{eq:partonYeik}.
Thus all the $Y$ dependence now occurs in the parton densities, while the function
$\widehat{\sigma}_0$ describes the interactions of two partons with the same rapidity.
In other words, in the alternative form \eqref{eq:partonYeik} there is no reference
to pomeron exchange, and all the Regge behaviour occurs in the densities --
the increase of the densities as a function of $y$ is natural because of the cascade
development of the two partonic systems.
The average number of slow partons, $\langle n_{i}(y)\rangle=\int d^2b_i~\rho_{i}(y,b_i)$,
is the product of the average number of fast partons, i.e. of partonic cascades, and
the average number of slow partons in the cascade, $m(y)\sim\exp(\Delta y)$.
However taking a Poisson distribution for the partons at each rapidity is
a strong assumption of the eikonal model.

Thus, in the framework of the parton model, we have either a Regge form of
$\widehat{\sigma}(Y)$ in (\ref{eq:partoneik}), or the Regge increase of the partonic densities
$\rho_{i}(y)$ in (\ref{eq:partonYeik}).
In (\ref{eq:partonYeik}), $\widehat{\sigma}_0$ describes a {\em local} parton-parton interaction
as a function of both rapidity and impact parameter, which avoids the highly non-local
pomeron interaction which occurs in (\ref{eq:partoneik}).
Energy-momentum conservation in the partonic interpretation can be imposed
by requirement of the energy-momentum sum rule for the parton distributions
$\rho_{i}$.

\subsection{Probabilistic interpretation of the inelastic cross section}

In the eikonal model there is a clear probabilistic interpretation of the inelastic
cross section \cite{aggk}.
The single inelastic cross section at fixed $b$, $\chi_P(Y,b)$,
corresponds to the product of the average numbers of partonic cascades at fixed $b_{1,2}$
and of probability for soft partons to interact. Similarly, the interaction of $k$ soft partons
from different parton chains is determined by the formula \eqref{eq:inel_k}
\begin{align*}
\sigma_k^{\inel}(Y,b) &= \frac{{(\chi_P(Y,b))}^k}{k!} \exp(-\chi_P(Y,b)) ~,
\end{align*}
where the exponential factor $\exp(-\chi_P)$ corresponds to the requirement that all other
partons do not interact.
The total inelastic cross section at fixed $b$ is, therefore,
\begin{align*}
\sigma^{\inel}(Y,b) = \sum_{k=1}^\infty \sigma_k^{\inel}(Y,b)
 &= 1 - \exp(-\chi_P(s,b) \nonumber\\
 &= \sum_{m=1}^\infty \frac{(-1)^{m-1}\chi_P^{m}(s,b)}{m!} ~.
\end{align*}
We can readily see the origin of the last, sign-alternating, expression for the
inelastic cross section. It is just a mathematical formula for the probability of joint
(not mutually excluded) events:
$$
\rm{Prob}(A_1\cup A_2\cup\dots\cup A_n) = \sum_{i}\rm{Prob}(A_i) -
\sum_{i<j}\rm{Prob}(A_i\cap A_j) + \sum_{i<j<k}\rm{Prob}(A_i\cap A_j\cap A_k)- \dots ~.
$$
An inelastic event corresponds to the interaction of at least one slow parton
with the target (an event ${\rm A_i}$). Then it is necessary to subtract from the formula for
$\sigma_1^{\inel}(Y,b)$ the probability of the interactions of two slow partons, and
to add the one for triple interaction, and so on.

It follows from the Poisson distribution (\ref{eq:inel_k})
that the average number, $\langle k \rangle$, of cut pomerons is equal to $\chi_P(Y,b)$.
It, thus, increases like $\exp(\Delta Y)$.
Thus, the increase of the produced particle densities, (\ref{eq:aincl}), in this model
is related to the very large number of partons in the hadronic fluctuations. For the
partonic interpretation \eqref{eq:partonYeik} with partonic cascading, the exponential increase
of the particle densities is clearly consistent with the inclusive particle formula
\eqref{eq:aincl}.

Thus, to interpret the case of $\Delta>0$, we have {\em either} to introduce
the energy dependence of the fast parton interactions (which is not self-consistent),
{\em or} to include a mechanism for splitting the partons in the course of the evolution in $y$.
In the latter case the number of slow partons increases as $\exp(\Delta Y)$, and if only one
of slow partons interacts with the target, then the exchange of the supercritical pomeron
is reproduced. The eikonal approximation will arise if several partons interact, but no more
than one parton of each independent (splitting) fluctuation (see Fig.4).
The more consistent approach, in which we allow any parton resulting from the fluctuation
to participate in the interaction, leads \cite{kgb} to the Schwimmer model and
its eikonalized version.
\begin{figure}[t]
\begin{center}
\includegraphics[width=0.8\textwidth]{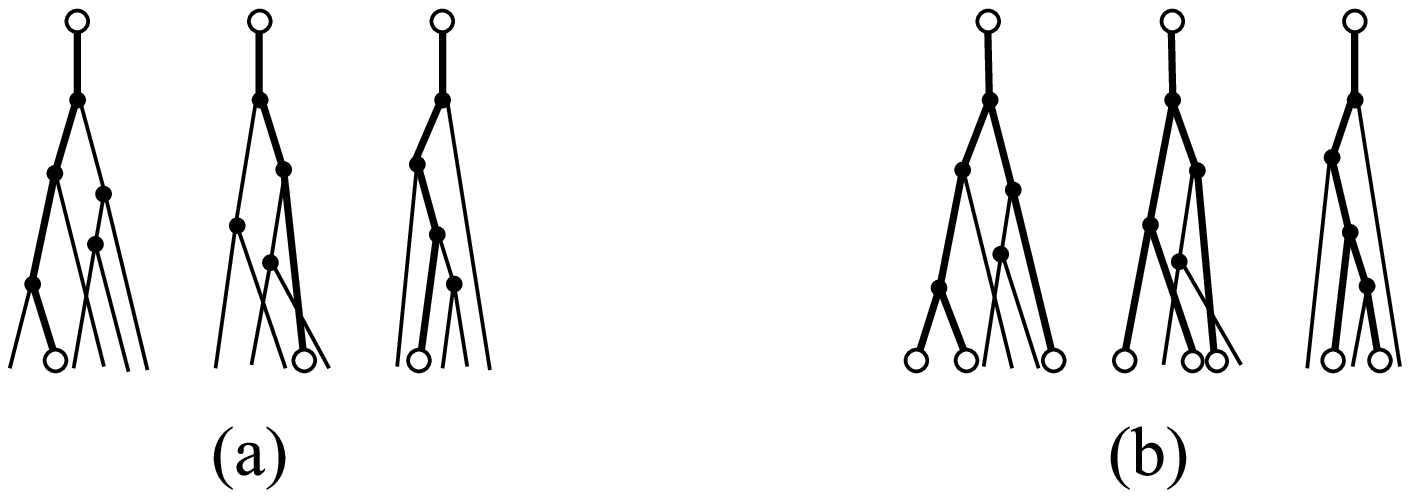}
\caption{
(a) Parton interpretation of the eikonal approximation for supercritical pomeron;
\mbox{(b) Parton} interpretation of the Schwimmer approximation (single cascade) and
of the eikonalized Schwimmer approximation (several cascades).
\label{fig:4}}
\end{center}
\end{figure}

\section{The Schwimmer model}

\begin{figure}[ht]
\begin{center}
\includegraphics[height=4cm]{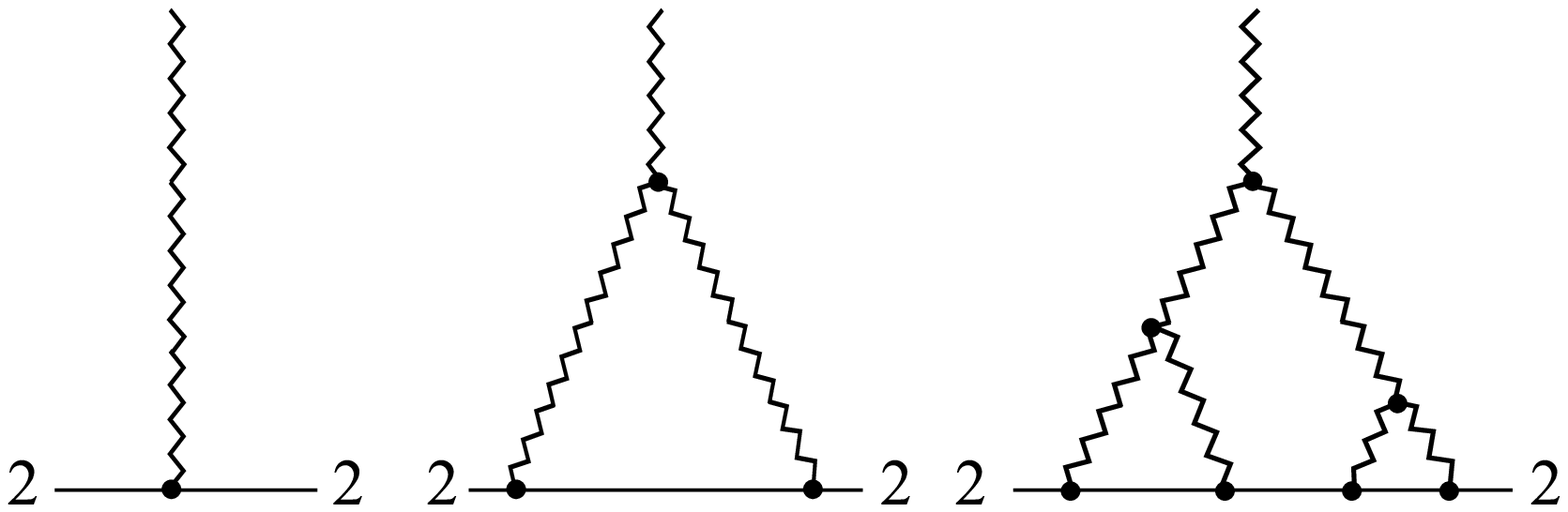}
\caption{Two typical fan diagrams orginating from the triple-pomeron coupling, which correct the original
pomeron exchange.\label{fig:5}}
\end{center}
\end{figure}

Schwimmer \cite{sch} proposed a simple model of reggeon field theory based on the
triple-pomeron interaction only. It sums up the set of fan diagrams of the type shown
in Fig.~\ref{fig:5}.
%

It is evident that the Schwimmer model is not symmetric with respect to the incoming
hadrons, 1 and 2. It was originally formulated for the interaction of hadrons with nuclei.
It is expected to be a reasonable approximation when hadron 1 (or a virtual photon)
has a size much smaller than hadron 2 (nucleus), that is $g_1 \ll g_2$, and the
triple-pomeron coupling $r$ is also small. In the case of the interaction with a
nucleus we may also neglect the dependence on the impact parameter, $b$, at energies
when the interaction radius is much smaller than the nuclear size.
Here we adopt this situation as a toy model
\footnote{~The introduction of the $b$ dependence is straightforward, but in this case
there is no analytic solution.}%
, so that the amplitudes depend only on the rapidity $Y$.

\subsection{Total cross section in the Schwimmer model}

\begin{figure}[ht]
\begin{center}
\includegraphics[width=12cm]{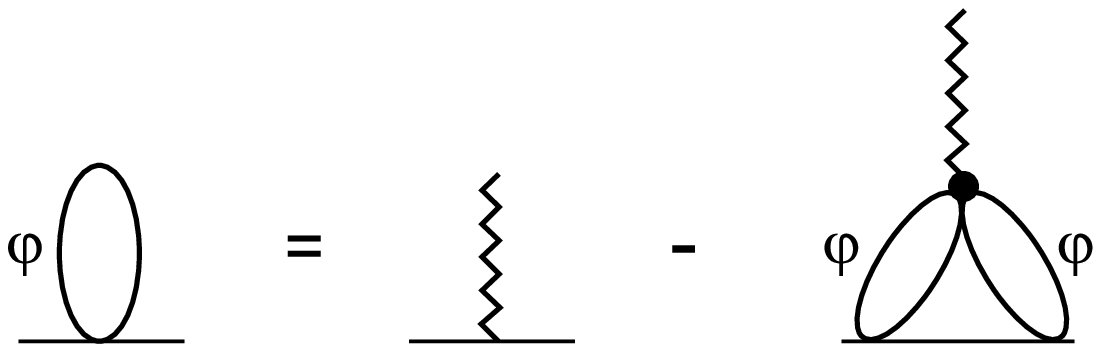}
\caption[Diagrammatic form]{Diagrammatic form of \eqref{eq:nonlin}, where here $\varphi$ denotes $\varphi_{\tot}/2$.
For clarity the diagram has not shown the truncation of particle 2.
\label{fig:6}}
\end{center}
\end{figure}
Following the Schwimmer model, we choose the $n$-pomeron-particle 2 amplitude to have
an eikonal form, with $G_n=g^n_2$. Rather than using the amplitude of \eqref{eq:eik},
it is convenient to work in terms of the `truncated' amplitude
$\varphi_{\tot}(Y)=\sigma^{\tot}(Y)/(g_1 g_2)= 2\Im f(Y)/(g_1 g_2)$,
and to introduce a new pomeron `propagator'
$P(Y)=\chi_P/(2g_1 g_2) = {\rm exp}(\Delta Y)$.
By construction, the function $\varphi_{\tot}(Y)$ satisfies the following non-linear
integral equation\footnote{Note that \eqref{eq:nonlin} is written for $\varphi_{\tot}(Y)/2$, since
the amplitude $f(Y)=ig_1g_2\varphi(Y)_{\tot}/2$.}
\begin{align} \varphi_{\tot}(Y)/2 = e^{\Delta Y} - r g_2
\int_0^Y dy_1 e^{\Delta(Y-y_1)}\,(\varphi_{\tot}(y_1)/2)^2~,
\label{eq:nonlin}
\end{align}
see Fig.~\ref{fig:6}. The differential form of the equation is
\begin{align}
\frac{d\varphi_{\tot}(Y)}{dY} = \Delta~\varphi_{\tot} - \frac{r g_2}{2} \varphi_{\tot}^2.
\label{eq:diffY}
\end{align}
To solve the equation it is convenient to make the substitution
\begin{align}
\varphi_{\tot}(Y) = 2\tau\; u_{\tot}(\tau), \qquad \tau = e^{\Delta Y} ~,
\label{eq:notation}
\end{align}
so that \eqref{eq:diffY} becomes
\begin{align}
\frac{d u_{\tot}(\tau)}{d\tau} = - \,\epsilon\, u_{\tot}^2 ~, \qquad  u_{\tot}(1) = 1 ~,
\qquad \text{with~} \epsilon = \frac{r g_2}{\Delta} ~.
\label{eq:tot}
\end{align}
The solution
\begin{align}
u_{\tot} = \frac{1}{1+\epsilon\,(\tau - 1)}
\label{eq:tot_sol}
\end{align}
gives the well known expression for $\varphi_{\tot}(Y)$
\begin{align}
\varphi_{\tot}(Y) = \frac{2 P(Y)}{1+\epsilon\, [P(Y)-P(0)]} ~.
\label{eq:sol}
\end{align}
Note that the integration in \eqref{eq:nonlin} goes from $y=0$ to $Y$. If, however,
the integration starts from $y_{\rm min}$, then for the corresponding solution
$\varphi_{\tot}(Y;y_{\rm min})$ we should replace $P(0)$ in \eqref{eq:sol} by $P(y_{\rm min})$.
From \eqref{eq:sol}, we see that the cross section $g_1 g_2 \varphi_{\tot}(Y)$ at first
increases as $\exp(\Delta Y)$, and then tends to the finite limit $2g_1\Delta/r$
for very large $Y$.

\subsection{Inelastic and diffractive cross sections}

To obtain the inelastic and diffractive amplitudes we use the AGK cutting rules
just as we did in section 3. We substitute for the cut amplitude the corresponding cross section
$\sigma^{\inel}$ or $\sigma^{\D}$, and for the uncut amplitude the factor $(-\sigma^{\tot})$.
This results in integral equations similar to
\eqref{eq:nonlin}, but with a non-diagonal structure for the inelastic cross section:
\begin{multline}
\label{eq:in}
\varphi_{\inel}(Y) \equiv \frac{\sigma^{\inel}}{g_1 g_2} =
 2 e^{\Delta Y}
 - 2 r\, g_2\int_0^{Y} dy_1 e^{\Delta(Y-y_1)} \varphi_{\inel}(y_1)\varphi_{\tot}(y_1) \\
  + {r\, g_2}\int_0^{Y} dy_1\, e^{\Delta(Y-y_1)} \varphi_{\inel}^2(y_1)
  + 2 r\, g_2 \int_0^{Y} dy_1\, e^{\Delta(Y-y_1)} \varphi_{\inel}(y_1) \varphi_{\D}(y_1)
 ~,
\end{multline}
while for the diffractive cross section, corresponding to the production of a state
accompanied by a rapidity gap,
\begin{multline}
 \varphi_{\D}(Y) \equiv \frac{\sigma^{\D}}{g_1 g_2} =
 \frac{r\, g_2}{2} \int_0^{Y} dy_1\, e^{\Delta(Y-y_1)}\varphi_{\tot}^2(y_1) \\
 - 2{r}\,{g_2}\int_0^{Y} dy_1
 e^{\Delta(Y-y_1)} \varphi_{\D}(y_1) \varphi_{\tot}(y_1)
 + {r\, g_2}\int_0^{Y} dy_1\, e^{\Delta(Y-y_1)} \varphi_{\D}^2(y_1) ~.
\label{eq:D}
\end{multline}
Note that coefficients in \eqref{eq:D}, which result from the different cuttings, are in the
same ratios, \mbox{($1: -4: 2$)}, as in~\eqref{cut2}.
Similar equations have been obtained in Ref. \cite{kovch99} in the framework of
the Balitsky-Kovchegov equation.

Taking into account that $\varphi_{\inel}+\varphi_{\D}=\varphi_{\tot}$, we obtain
from \eqref{eq:in} the differential equation for inelastic cross section
\begin{alignat}{2}
\frac{d u_{\inel}(\tau)}{d\tau} &= -2 \epsilon\, u_{\inel}^2 ~,
&\quad & u_{\inel}(1) = 1 ~,
\label{eq:diff}
\end{alignat}
where, similar to {\eqref{eq:notation}, we use the substitutions
\begin{align*}
\varphi_{\inel}(Y) = 2\tau\, u_{\inel}(\tau), \qquad \varphi_{\D}(Y) = 2\tau\, u_{\D}(\tau) ~.
\end{align*}
It is remarkable that the equation for $u_{\inel}$, i.e. for $\sigma^{\inel}$, is
diagonal. It is a generalization of the similar result for the non-enhanced diagrams
(see the footnote in section \ref{f1}).

Thus, in analogy to \eqref{eq:sol}, we obtain the solutions
\begin{align}
 \label{sig_inel}
 u_{\inel}(\tau) &= \frac{1}{1+2\epsilon\,(\tau-1)} ~, \\
 \label{sig_D}
 u_{\D}(\tau) &\equiv u_{\tot}(\tau) - u_{\inel}(\tau)
 = \frac{1}{1+\epsilon\,(\tau-1)} - \frac{1}{1+2\epsilon\,(\tau-1)} ~.
\end{align}
Note that in the limit $\epsilon\tau \ll 1$, these solutions
reproduce the first reggeon graphs, and that in the saturation regime
(where $\epsilon\tau \gg 1$) we have
$\varphi_{\inel}=\varphi_{\rm D}=\varphi_{\tot} /2=1/\epsilon$.


Next, we obtain the dependence of diffractive production on the rapidity gap $y$, or
on the mass of the produced system, where ln$(M^2/M_0^2)=Y-y$.
We introduce a function $\varphi_{\rm gap}(Y;y_{\rm min})$ corresponding
to the cross section for production of the final state with a rapidity gap
larger than $y_{\rm min}$:
\begin{align}
\varphi_{\rm gap}(Y;y_{\rm min}) = \frac{1}{g_1 g_2}\int_{y_{\rm min}}^Y\! dy_1
\;\frac{d\sigma^{\rm D}}{dy_1} ~, \qquad \varphi_{\rm gap}(Y;0)=\varphi_{\rm D} ~.
\label{eq:def_gap}
\end{align}
This cross section satisfies the same integral equations as the diffractive dissociation
cross section $\varphi_{\rm D}$ except that the integration over rapidity
starts from $y_{\rm min}$ instead of 0. That is
\begin{align}
\varphi_{\rm gap}(Y;y_{\rm min}) &=
\frac{rg_2}{2}\int_{y_{\rm min}}^{Y} e^{\Delta(Y-y_1)} \varphi_{\rm tot}^2(y_1)
 \nonumber\\
& \qquad
- 2 {r}{g_2}\int_{y_{\rm min}}^{Y} dy_1 e^{\Delta(Y-y_1)}
\varphi_{\rm gap}(y_1;y_{\rm min}) \varphi_{\tot}(y_1)  \nonumber\\
& \phantom{quad - 4 {r}{g_2}\int_{y_{\rm min}}^{Y} dy_1 e^{\Delta(Y-y_1)}\varphi_{\rm gap}}
\quad +
 {r}{g_2}\int_{y_{\rm min}}^{Y} dy_1\, e^{\Delta(Y-y_1)} \varphi_{\gap}^2(y_1;y_{\rm min})
 ~,
\label{eq:gap}
\end{align}
see Fig.~\ref{fig:7}.

\begin{figure}
\begin{center}
\includegraphics[width=0.9\textwidth]{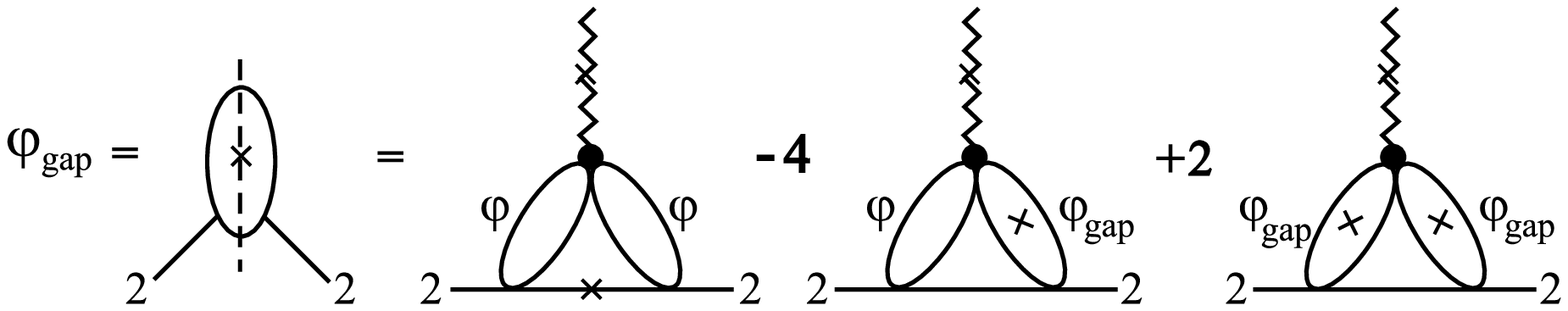}
\caption{Diagrammatic form of \eqref{eq:gap}, where $\varphi$ denotes $\varphi_{\tot}$.
For clarity the diagram has not shown the truncation of particle 2. Recall that the extra factor of 1/2 in \eqref{eq:gap} in the ratios \mbox{($1: -4: 2$)} is because the amplitude $f(Y) \propto \varphi/2$.
\label{fig:7}}
\end{center}
\end{figure}
As before, we may write this in the differential form
\begin{alignat}{2}
\frac{d u_{\rm gap}(\tau;\tau_{\rm min})}{d\tau} =
 2\epsilon\, (\frac{1}{2} u_{\tot}^2 - 2 u_{\ gap} u_{\tot} + u_{\gap}^2) ~,
 \qquad  u_{\rm gap}(\tau_{\rm min};\tau_{\rm min})=0 ~,
\label{eq:diff_gap}
\end{alignat}
where $u_{\tot}$ is the solution of \eqref{eq:tot}, and
\begin{align}
 \varphi_{\rm gap}(Y;y_{\rm min}) = 2\tau\; u_{\rm gap}(\tau;\tau_{\rm min}) ~, \quad 
 \tau_{\rm min} = e^{\Delta y_{\rm min}} ~.
\label{eq:notation_gap}
\end{align}
In analogy to \eqref{sig_D}, the solution is
\begin{align}
 \label{sig_DD}
 u_{\rm gap}(\tau;\tau_{\rm min}) &
 = \frac{1}{1+\epsilon(\tau-1)} - \frac{1}{1+\epsilon\,(2\tau-\tau_{\rm min}-1)} ~,
\end{align}
or,
\begin{align}
 \label{sig_DDD}
 \varphi_{\rm gap}(Y;y_{\rm min}) &= \frac{2e^{\Delta Y}}
 {1+\epsilon\,(e^{\Delta Y}-1)}
  - \frac{2e^{\Delta Y}}
 {1+\epsilon\,(2e^{\Delta Y}-e^{\Delta y_{\rm min}}-1)} ~.
\end{align}
Thus, we can calculate the cross section for a fixed gap $y$, that is for the
diffractive production of a state of given mass $M$ (with the value of $y_M = Y - y$ fixed).
It is determined by the derivative of the second term of \eqref{sig_DDD}:
\begin{align}
\label{eq:sig_M_sch}
\frac{d\sigma^{\rm D}}{dy_M} \equiv M^2\frac{d\sigma^{\rm D}}{dM^2}
&= - g_1 g_2 \frac{d\varphi_{\rm gap}(Y;y)}{dy}
= \frac{2 g_1 g_2\Delta\epsilon\,e^{\Delta (2 Y - y_M)}}
 {[1+\epsilon\,(2e^{\Delta Y}-e^{\Delta (Y - y_M)}-1)]^2} \\[2mm]
 \label{eq:limit}
 &\approx
\frac{g_1\Delta^2}{r}\frac{2\exp (\Delta y_M)}{[2\exp(\Delta y_M)-1]^2}
\quad \text{(~for $\epsilon \exp(\Delta y_{\min})\gg 1$~)} ~.
\end{align}
This cross section in the Schwimmer model was first obtained in Ref.~\cite{bond00}.

We see that the cross section \eqref{eq:limit} decreases with $M^2$, which provides
convergence of the integral over the mass of the diffractively produced system.
Indeed, in the region of large $M^2$, that is in the saturation domain with $y_M \gg 1$,
we have
\be
M^2\frac{d\sigma^{\rm D}}{dM^2} \sim (M^2)^{-\Delta}
\label{eq:mm}
\ee
Thus the $M^2$ distribution gives information on the intercept of the bare pomeron,
$\alpha_P(0) \equiv 1+\Delta$.
Although \eqref{eq:mm} was derived in the Schwimmer model,
we shall see that the same behaviour is valid for its eikonal generalization.

Another way to get information on the bare intercept is to study the inclusive spectrum.
Using the AGK cutting rules, we find that the particle rapidity distribution is
\begin{align}
\frac{d\sigma^a}{dy} = \lambda_a\, g_1 g_2\,e^{\Delta y}\,\varphi_{\tot}(y_2) ~,
\qquad {\rm with} \quad y_2=Y-y.
\label{eq:rap}
\end{align}

In a frame where hadron 1 is moving fast, \eqref{eq:rap} can be interpreted as
a Regge-like increase of partons.  However, the partonic interpretation
of this result is different in a frame where particle 2 is fast;  see section 4.4.

\subsection{The eikonalized Schwimmer model}

Suppose, now, that there are several partons in the initial state at $y=0$ which split
in the course of the evolution. In the absence
of splitting this would correspond to the usual eikonal model
(see sect. 3). However as a result of splitting, the evolution of each initial parton
corresponds to the Schwimmer amplitude -- and the whole amplitude is described by Fig.~\ref{fig:8}.
The AGK rules for this set of diagrams are similar to the ones for the eikonal graphs of Fig.1
except that each Schwimmer-type amplitude contains, not only the inelastic discontinuity
$\varphi_{\inel}$ due to pomeron exchange, but also the discontinuity corresponding to
gap production $\varphi_{\gap}(Y;y_{\min})$, with relations
$\varphi_{\inel}+\varphi_{\D}=\varphi$ and $\varphi_{\D}=\varphi_{\gap}(Y;0)$.
Then, the set of formulae for the various cross sections will be similar to the
\eqref{eik_tot}, \eqref{eik_inel}, \eqref{eik_diff},
together with the ones resulting from the extra discontinuities of the amplitude
\begin{align}
\label{sigma_tot}
\sigma^{\tot}(Y;b) &= 2\left[ 1-\exp(- g_1 g_2 \varphi_{\tot}(Y)/2)\right] ~, \\[2mm]
\label{sigma_el}
\sigma^{\el}(Y;b) &= \left( 1-\exp(- g_1 g_2 \varphi_{\tot}(Y)/2)\right)^2 ~, \\[2mm]
\label{sigma_inel}
\sigma^{\inel}(Y;b) &= 1 - \exp(-g_1 g_2 \varphi_{\inel}(Y)) ~, \\[2mm]
\label{sigma_D}
\sigma^{\D}(Y;b) &= \exp(-g_1 g_2 \varphi_{\inel}(Y)) - \exp(-g_1 g_2 \varphi_{\tot}(Y)) ~, \\[2mm]
\label{sigma_gap}
\sigma^{\gap}(Y;y_{\min};b) &= \exp(-g_1 g_2 \varphi_{\tot}(Y))
\left[(\exp(g_1 g_2 \varphi_{\gap}(Y;y_{\min})) - 1\right] ~,
\end{align}
where $\varphi_{\tot}$, $\varphi_{\inel}$, $\varphi_{\D}$ and  $\varphi_{\gap}$ have been defined
above. We see that the following relations hold
\begin{align}
\label{sigma_tot_el}
\sigma^{\tot}(b) - \sigma^{\el}(b) = \sigma^{\inel}(b) + \sigma^{\D}(b)
 = 1 - e^{-g_1 g_2 \varphi_{\tot}} ~.
\end{align}
Note that again we have a closed expression for $\sigma^{\inel}$.
\begin{figure}
\begin{center}
\includegraphics[height=5cm]{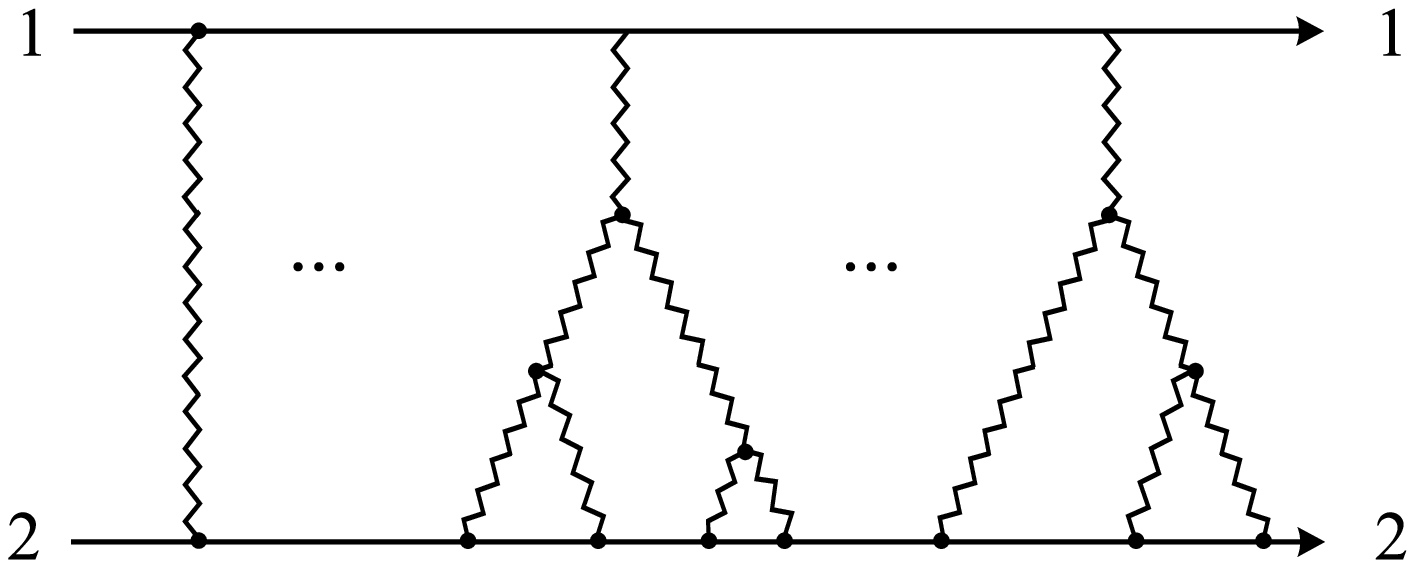}
\caption{The eikonalized version of the Schwimmer model,
which is a generalization of~Figs.1(a),~5.
\label{fig:8}}
\end{center}
\end{figure}

The differential cross section for the diffractive production of a state of mass $M$
is obtained by differentiation of \eqref{sigma_gap} with respect to $y_{\min}$, which
enters via $\varphi_{\gap}$. That is via
\begin{align}
\label{eq:sig_M}
\frac{d\sigma^{\gap}}{dy_M} =
\exp\left[-g_1 g_2 (\varphi_{\tot}(Y;b) - \varphi_{\gap}(Y,Y-y_M;b))\right]
\frac{d\sigma_{\rm Sch}^{\rm D}}{dy_M}
 ~,
\end{align}
where $d\sigma_{\rm Sch}^{\rm D}/dy_M = - g_1 g_2 d\varphi_{\rm gap}(Y;y)/dy$ is defined by
\eqref{eq:sig_M_sch}.
In the saturation limit \eqref{eq:limit} we obtain
\begin{align}
\frac{d\sigma^{\rm D}}{dy_M}\approx
\exp\left(-\frac{g_1\Delta}{r}\right)
\frac{g_1\Delta^2}{r}\frac{2\exp (\Delta y_M)}{[2\exp(\Delta y_M)-1]^2} ~,
\end{align}
for $1\ll y_M\ll Y$. Thus, again, the dependence shown in \eqref{eq:mm} is valid
at large values of $M$.

We note that equations \eqref{sigma_D}, \eqref{sigma_gap}, \eqref{eq:sig_M}
differ from the results of \cite{bond00}, where absorptive effects were included by multiplication by the factor $\exp(-\varphi_{\rm tot})$. This procedure,
however, does not allow for the simultaneous diffractive cuttings of several Schwimmer amplitudes.
This difference is especially important in calculations of the survival probability which take
into account absorptive effects in inelastic diffractive processes.
For example, for the inclusive production of particles in large-mass diffraction,
the survival probability has the form
\begin{align}
\label{surv_prob}
S^2(Y,y_M;b)=
\exp\left[-g_1 g_2 (\varphi_{\tot}(Y;b) - \varphi_{\gap}(Y,Y-y_M;b))\right] ~.
\end{align}
This result can be easily obtained by using the method of Ref. \cite{bkai}. We emphasize that,
in contrast to the eikonal model, the survival probability depends not only on $Y$, but also
on the mass of the produced system $y_M$.


\subsection{Partonic interpretation of Schwimmer diagrams}

As we discussed in \ref{partons_eik}, the supercritical pomeron
requires a mechanism for parton splitting. In the Schwimmer model this occurs
through a single parton cascading in terms of reggeon diagrams.
On the other hand, in the eikonalized Schwimmer model it is described by the independent
cascading of a Poisson set of initial fast partons. In both models the inclusive spectrum
is described by similar formula \eqref{eq:rap}. The increase of the spectrum with the rapidity
of the inclusive particle is due to the partonic cascade, which leads
to an exponential growth of partons with $y$. Note that there is no fusion of partons
in this cascade, which would have inhibited its growth.

We stress that the model is not symmetric with respect to the colliding hadrons.
In the frame where hadron 2 moves fast, the parton interpretation requires both
splitting and fusion of partons like the first and the second terms in the r.h.s. of
Eq.\eqref{eq:diffY}). As a result, we first have a growth of the number of partons
and then saturation to a constant value, due to recombination.
This is the usual interpretation used in discussions of the saturation of parton
densities in QCD cascade \cite{bk}.
This behaviour of parton density, in the case where hadron 2 moves fast,
can be traced to the $y_2$ behaviour of the inclusive spectrum \eqref{eq:rap}.
Note, however, that only parton fusion producing tree reggeon diagrams is allowed
in this approximation. This is justified for $r\ll 1, g_2\gg 1$, as was discussed in section 4.

Note that this dependence of the partonic interpretation on the choice of
the Lorentz frame is due to the special (non-symmetric) selection of reggeon diagrams
related to particular process. This is reasonable in a limited region of rapidity,
with $r\exp(\Delta Y)\ll 1$. For higher rapidities, loop diagrams become important.
Of course, if the complete set of diagrams of reggeon theory were to be used, then
the parton dynamics would be identical in all Lorentz frames \cite{kgb}.

The multiplicity distribution in the Schwimmer model is not Poisson-like \cite{kgb}.
There are huge fluctuations, leading, at high energies, to a large dispersion.
Hence, according to the Good-Walker formalism \cite{gw}, there is a large probability
of diffractive dissociation.

\section{Conclusions}

We have investigated the two simplest models of reggeon theory, using the AGK cutting rules
for the supercritical pomeron. We discussed the partonic interpretation of the models.
A closed set of equations is obtained for $\sigma^{\tot}$, $\sigma^{\inel}$ and $\sigma^{\D}$
in the Schwimmer model. It is important that the equation for $\sigma^{\inel}$ is diagonal,
as is the equation for $\sigma^{\tot}$. Explicit formulae for rapidity gap
production are obtained.

We note that, from the partonic viewpoint, both the eikonal and Schwimmer models
are incomplete at asymptotic energies. In particular, the partonic interpretation
of the Schwimmer model depends on the choice of the Lorentz frame.
We note that, at asymptotic energies, partonic dynamics must be Lorentz invariant. From the viewpoint of reggeon field theory, this corresponds to the crucial
role of the pomeron loops.

The extension of the multi-pomeron formalism carried out in this paper can lead to a better
understanding of high energy dynamics and to an improvement of the analysis of data
for soft high energy interactions.
This is important, for example, in the calculation of probabilities of rapidity gaps
in diffractive processes; see, for example, Ref.~\cite{KMRsoft}.

\section*{Acknowledgements}

We thank E. Levin for drawing our attention to Refs.~\cite{kovch99,bond00}, and
for useful discussions.
ABK and MGR would like to thank the IPPP at the University of Durham for hospitality,
and ADM thanks the Leverhulme Trust for an Emeritus Fellowship. This work was supported
by the Royal Society, the UK Particle Physics and Astronomy Research Council,
by grants CRDF RUP2-2621-MO-04, RFBR 04-02-16073, 04-02-17263 and 03-02-04004,
SS-1124.2003.2 and SS-1774.2003.2.



\end{document}